\documentclass{paper}
\usepackage{amsfonts,amsmath,latexsym,amscd,graphicx,amssymb}

\newtheorem{theorem}{Theorem}

\newcommand{\f}{\frac}

\newcommand{\be}{\begin{equation}}
\newcommand{\ee}{\end{equation}}
\newcommand{\ba}{\begin{eqnarray}}
\newcommand{\ea}{\end{eqnarray}}

 \title{Particle trajectories in linearized irrotational shallow water
flows }

\author{\normalsize Delia IONESCU-KRUSE\\
\normalsize Institute of Mathematics of
the Romanian Academy,\\
\normalsize P.O. Box 1-764, RO-014700, Bucharest,
 Romania\\
\normalsize E-mail: Delia.Ionescu@imar.ro\\[10pt]}

 \date{}

\begin{document}

\maketitle
\begin{abstract}
\noindent
\end{abstract}
We investigate the particle trajectories in an irrotational
shallow water flow over a flat bed as  periodic waves propagate on
the water's free surface. Within the linear water wave theory, we
show that there are no closed orbits for the  water particles
beneath the irrotational shallow water waves. Depending on the
strength of underlying uniform current,  we obtain that some
particle trajectories are undulating path to the right or to the
left, some are looping curves with a drift to the right and others
are  parabolic curves or curves which have only one loop.

\section{Introduction}

The motion of water particles  under the waves which advance
across the water is a very old problem. The classical description
of these particle paths is obtained within the framework of linear
water wave theory.  After the linearization of the governing
equations for water waves, the ordinary differential equations
system which describes the particle motion turns out to be again
nonlinear and explicit solutions are not available. In the first
approximation of this nonlinear system, one obtained that all
water particles trace closed, circular or elliptic, orbits (see,
for example, \cite{debnath}, \cite{johnson-carte}, \cite{lamb},
\cite{lighthill}, \cite{sommerfeld}, \cite{stoker}),
 a conclusion apparently supported  by photographs with long exposure
 (\cite{debnath}, \cite{sommerfeld}, \cite{stoker}).
Consequently there is no net transfer of material particles due to
the passage of the wave, at least, at this order of
approximation.\\
While in these approximations of the nonlinear system all particle
paths appear to be closed, in \cite{cv} it is shown, using
phase-plane considerations, that in linear periodic gravity water
waves no particles trajectory is actually closed, unless the free
surface is flat. Each particle trajectory involves over a period a
backward/forward movement, and the path is an elliptical arc with
a forward drift; on the flat bed the particle path
degenerates to a backward/forward motion. \\
Similar results hold for the particle trajectories in deep-water,
that is, the trajectories are not closed existing  a forward drift
over a period, which decreases with greater depth (see
\cite{cev}). These conclusions are in agreement with Stokes'
observation \cite{stokes}: "There is one result of a second
approximation which may possible importance. It appears that the
forward motion of the particles is not altogether compensated by
their backward motion; so that, in addition to their motion of
oscillation, the particles have a progressive motion in the
direction of the propagation of the waves. In the case in which
the depth of the fluid is very great, this progressive motion
decreases rapidly as the depth of the particle considered
increases."\\
For shallow water waves, the standard results are that the orbits
described by water particles beneath waves are elongated ellipses
with the longer axis parallel to the flat bottom, and at the
bottom the orbits are straight lines (see, for example,
\cite{kk}).

Similar conclusions hold for the governing equations without
linearization. Analyzing a free boundary problem for harmonic
functions in a planar domain, in \cite{c2007} it is shown that
there are no closed orbits for Stokes waves of small or large
amplitude propagating at the surface of water over a flat bed; for
an extension of the investigation in \cite{c2007} to deep-water
Stokes waves see \cite{henry}. Within a period each particle
experiences a backward/forward motion with a slight forward drift.
In a very recent preprint \cite{CS},  the results in \cite{c2007}
are recovered by a simpler approach and there are also described
all possible particle trajectories beneath a Stokes wave. The
particle trajectories change considerably according to whether the
Stokes waves enter a still region of water or whether they
interact with a favorable or adverse uniform current. Some
particle trajectories are closed orbits, some are undulating paths
and most are looping orbits that drift either to the right or to
the left, depending
on the underlying current. \\
Analyzing a free boundary problem for harmonic functions in an
infinite planar domain, in \cite{CE2} it is shown that under a
solitary wave, each particle is transported in the wave direction
but slower than the wave speed. As the solitary wave propagates,
all particles located ahead of the wave crest are lifted while
those behind have a downward motion.

 Notice that there are only a
few explicit solutions to the nonlinear governing equations:
Gerstner's wave (see \cite{gerstner} and the discussion in
\cite{c2001a})), the edge wave solution related to it (see
\cite{c2001b}), and the capillary waves in water of infinite or
finite depth (see \cite{crapper}, \cite{kinn}). These solutions
are peculiar and their special features (a specific vorticity for
Gerstner's wave and its edge wave correspondent, and complete
neglect of gravity in the capillary case) are not deemed relevant
to sea waves.

The present paper is concerned with  the particle trajectories in
an irrotational shallow water flow over a flat bed as a periodic
wave propagates on the water's free surface. It is natural to
start this investigation for shallow water waves by simplifying
the governing equations via linearization. In  Section 2 we recall
the governing equations for water waves. In Section 3 we present
their nondimensionalisation and scaling. The linearized problem in
the irrotational shallow water regime is written in Section 4. We
also obtain the general solution of this problem. The next section
is devoted to the description of all the possible particle
trajectories beneath a linear periodic irrotational shallow water
wave. We see that these particle trajectories are not closed.
Depending on the strength of underlying uniform current, denoted
by the constant $c_0$, we obtain that: for $c_0>2$ the particle
trajectories are undulating path to the right, for $c_0<-1$ the
particle trajectories are undulating path to the left, for $-1\leq
c_0<0$ the particle trajectories are looping curves with a drift
to the right and for $0\leq c_0\leq 2$ the particle trajectories
are parabolic curves or curves which have only one loop.

\section{The governing
equations for gravity
 water waves}

 We consider a two-dimensional inviscid incompressible fluid in a
constant gravitational field. For gravity water waves these are
physically reasonable assumptions (see \cite{johnson-carte} and
\cite{lighthill}). Thus, the motion of water is given by Euler's
equations
\begin{equation}
\begin{array}{c}
u_t+uu_x+vu_z=-\f1{\rho} p_x\\  v_t+uv_x+vv_z=-\f1{\rho} p_z-g\\
\end{array}
 \label{e}
 \end{equation}
Here $(x,z)$ are the space coordinates, $(u(x,z,t), v(x,z,t))$ is
the velocity field of the water, $p(x,z,t)$ denotes the pressure,
$g$ is the constant gravitational acceleration in the negative $z$
direction and  $\rho$ is the constant density. The assumption of
incompressibility implies the equation of mass conservation \be
u_x+v_z=0 \label{mc}\ee

\noindent  Let $h_0>0$ be the undisturbed depth of the fluid and
let $z=h_0+\eta(x,t)$ represent the free upper surface of the
fluid (see Figure 1).
\\

 \hspace{2cm}\scalebox{0.65}{\includegraphics{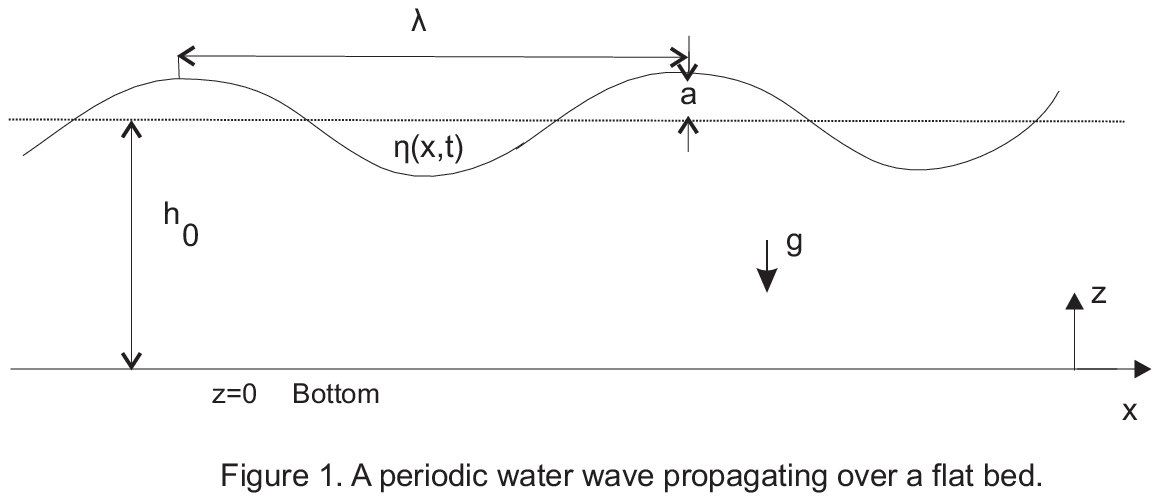}}
\\

\noindent
 The boundary conditions at the free surface are
constant pressure
\begin{equation}
p=p_0 \,  \textrm{ on } z=h_0+\eta(x,t),
 \label{bc2}
 \end{equation}
 $p_0$ being the constant
atmospheric pressure, and the continuity of fluid velocity and
surface velocity
\begin{equation}
  v=\eta_t+u\eta_x \, \, \textrm{ on }\,
z=h_0+\eta(x,t)
 \label{bc1}
 \end{equation}
On the flat bottom $z=0$, only one condition is required for an
inviscid fluid, that is,
\begin{equation}
 v=0 \, \,
\textrm { on } z=0
 \label{bc1'}
 \end{equation}
Summing up, the exact  water-wave problem is given by the system
(\ref{e})-(\ref{bc1'}). In respect of the well-posedness for the
initial-value problem for (\ref{e})-(\ref{bc1'}) there has been
significant recent progress, see \cite{shkoller} and the
references therein.

A key quantity in fluid dynamics is   the \emph{curl} of the
velocity field, called vorticity. For two-dimensional flows with
the velocity field $(u(x,z,t), v(x,z,t))$, we denote the scalar
vorticity of the flow by \be \omega(x,z)=u_z-v_x
\label{vorticity}\ee Vorticity is adequate for the specification
of a flow:
 a flow which is uniform
 with depth is described by a zero vorticity (irrotational case),
 constant non-zero
 vorticity corresponds to a linear shear flow and non-constant vorticity
 indicates highly sheared flows.

 The full Euler equations (\ref{e})-(\ref{bc1'}) are often too complicated
 to analyze directly.
One can pursue for example a mathematical study of their periodic
steady solutions
 in the irrotational case (see \cite{aft}, \cite{toland}) or
 a study of their  periodic steady
solutions in the case of non-zero vorticity (see \cite{cs2'},
\cite{cs2007}). But in order to reach detailed information about
qualitative features of water waves, it is useful to derive
approximate models which are more amenable to an in-depth
analysis.

\section{Nondimensionalisation and scaling}

In order to develop a systematic approximation procedure, we need
to characterize the water-wave problem (\ref{e})-(\ref{bc1'}) in
terms of the sizes of various fundamental parameters. These
parameters are introduced by defining a set of non-dimensional
variables.\\
First we introduce the appropriate length scales: the undisturbed
depth of water $h_0$, as the vertical scale and a typical
wavelength $\lambda$ (see Figure 1), as the horizontal scale. In
order to define a time scale we require a suitable velocity scale.
An appropriate choice for the scale of the horizontal component of
the velocity is $\sqrt{gh_0}$. Then, the corresponding time scale
is $\f\lambda{\sqrt{gh_0}}$ and the scale for the vertical
component of the velocity is $h_0\f{\sqrt{gh_0}}{\lambda}$. The
surface wave itself leads to the introduction of a typical
amplitude of the wave $a$ (see Figure 1). For more details see
\cite{johnson-carte}. Thus, we define the set of non-dimensional
variables
\begin{equation}
\begin{array}{c}
x\mapsto\lambda x,  \quad z\mapsto h_0 z, \quad \eta\mapsto a\eta,
\quad t\mapsto\f\lambda{\sqrt{gh_0}}t,\\
  u\mapsto  \sqrt{gh_0}u,
\quad v\mapsto h_0\f{\sqrt{gh_0}}{\lambda}v
\end{array} \label{nondim}\end{equation}
where, to avoid new notations, we have used the same symbols for
the non-dimensional variables  $x$, $z$, $\eta$, $t$, $u$, $v$, on
the right-hand side. The partial derivatives will be replaced by
\begin{equation}
\begin{array}{c}
u_t\mapsto \f{gh_0}{\lambda}u_t, \quad u_x\mapsto
\f{\sqrt{gh_0}}{\lambda}u_x, \quad u_z\mapsto\f {\sqrt{gh_0}}{h_0}u_z,\\

v_t\mapsto \f{gh_0^2}{\lambda^2}v_t, \quad v_x\mapsto
h_0\f{\sqrt{gh_0}}{\lambda^2}v_x, \quad v_z\mapsto\f {\sqrt{gh_0}}{\lambda}v_z\\
\end{array}\label{derivate}\end{equation}

\noindent Let us now define the non-dimensional pressure. If the
water would be stationary, that is, $u\equiv v \equiv 0$, from the
equations (\ref{e}) and (\ref{bc2}) with $\eta=0$, we get for a
non-dimensionalised $z$, the hydrostatic pressure $p_0+\rho g
h_0(1-z)$. Thus, the non-dimensional  pressure is defined  by
\begin{equation} p\mapsto p_0+\rho g h_0(1-z)+\rho g h_0 p
\label{p}\end{equation} therefore
\begin{equation} p_x\mapsto \rho \f {gh_0}{\lambda} p_x, \quad
p_z\mapsto -\rho g+\rho g p_z\label{p'}\end{equation}

 Taking
into account (\ref{nondim}), (\ref{derivate}), (\ref{p}) and
(\ref{p'}), the water-wave problem (\ref{e})-(\ref{bc1'})  writes
 in
non-dimensional variables, as \begin{equation}
\begin{array}{c}
u_t+uu_x+vu_z=- p_x\\  \delta^2(v_t+uv_x+vv_z)=- p_z\\
 u_x+v_z=0\\
v=\epsilon(\eta_t+u\eta_x) \,   \textrm{ and } \,  p=\epsilon\eta
\, \, \textrm{ on }\,
z=1+\epsilon\eta(x,t)\\
 v=0 \, \,
\textrm { on } z=0
 \end{array}
\label{e+bc'} \end{equation}  where we have introduced the
amplitude parameter $\epsilon=\f a{h_0}$ and the shallowness
parameter $\delta=\f {h_0}{\lambda}$. In view of (\ref{derivate}),
the
 vorticity equation (\ref{vorticity}) writes in non-dimensional variables
 as
 \be
 u_z=\delta^2v_x+\f{\sqrt{gh_0}}{g}\omega(x,z)\label{vor}
 \ee
For zero vorticity flows (irrotational flows) this equation writes
as
 \be
 u_z=\delta^2v_x\label{vor1}
 \ee

After the nondimensionalisation of  the system
(\ref{e})-(\ref{bc1'}) let us now proceed  with the scaling
transformation. First we observe that, on $z=1+\epsilon\eta$, both
$v$ and $p$ are proportional to $\epsilon$. This is consistent
with the fact that as $\epsilon\rightarrow 0$ we must have
$v\rightarrow 0$ and $p\rightarrow 0$, and it leads to the
following scaling of the non-dimensional variables
\begin{equation} p\mapsto \epsilon p,\quad
(u,v)\mapsto\epsilon(u,v) \label{scaling}\end{equation} where we
avoided again the introduction of a new notation. The problem
(\ref{e+bc'}) becomes \begin{equation}
\begin{array}{c}
u_t+\epsilon(uu_x+vu_z)=- p_x\\  \delta^2[v_t+\epsilon(uv_x+vv_z)]=- p_z\\
 u_x+v_z=0\\
  v=\eta_t+\epsilon u\eta_x  \, \textrm{ and } \,  p=\eta \, \, \textrm{ on }\,
z=1+\epsilon\eta(x,t)\\
 v=0 \, \,
\textrm { on } z=0
 \end{array}
\label{e+bc1''} \end{equation} and the equation (\ref{vor}) keeps the same form. \\
 The system which
describes our problem  in the irrotational case is given by \be
\begin{array}{c}
u_t+\epsilon(uu_x+vu_z)=- p_x\\  \delta^2[v_t+\epsilon(uv_x+vv_z)]=- p_z\\
 u_x+v_z=0\\
  u_z=\delta^2v_x\\
  v=\eta_t+\epsilon u\eta_x  \, \textrm{ and } \,  p=\eta \, \, \textrm{ on }\,
z=1+\epsilon\eta(x,t)\\
 v=0 \, \,
\textrm { on } z=0
 \end{array}
\label{e+bc''} \end{equation}

\section{The linearized problem}

The two important  parameters $\epsilon$ and $\delta$  that arise
in water-waves theories, are used to define various approximations
of the governing equations and the boundary conditions. The scaled
version (\ref{e+bc''})  of the equations for our problem, allows
immediately the identification of the linearized problem, by
letting $\epsilon\rightarrow 0$, for arbitrary $\delta$.  The
linearized problem in the shallow water regime is obtain by
letting further $\delta\rightarrow 0$. Thus, in the irrotational
case, we get  the following linear systems
\begin{equation}
\begin{array}{c}
u_t+p_x=0\\  p_z=0\\
 u_x+v_z=0\\
 u_z=0\\
v=\eta_t \,  \textrm{ and } \,  p=\eta \, \, \textrm{ on }\,
z=1\\
 v=0 \, \,
\textrm { on } z=0
\end{array}
\label{small} \end{equation} From the second equation in
(\ref{small}) we get in the both cases that $p$ does not depend on
$z$. Because $p=\eta(x,t)$ on $z=1$, we have
\begin{equation} p=\eta(x,t) \, \quad \textrm{ for any } \, \,
0\leq z\leq 1\label{2}\end{equation} Therefore, using the first
equation and the fourth equation in (\ref{small}),  we obtain in
the irrotational case
 \begin{equation} u=-\int_0^t \eta_x(x,s)ds+\mathcal{F}(x)
\label{3''}\end{equation} where $\mathcal{F}$ is an arbitrary
function such that \be
 \mathcal{F}(x)=u(x,0)
\ee
\\
 Differentiating
(\ref{3''}) with respect to $x$ and using the third equation in
(\ref{small}) we get, after an integration against $z$,
 \be
v=-zu_x=z\left(\int_0^t\eta_{xx}(x,s)ds
-\mathcal{F}'(x)\right)\label{4}\ee In view of the fifth equation
in (\ref{small}) we get after a differentiation with respect to
$t$, that $\eta$ has to satisfy the equation
 \begin{equation}
\eta_{tt}-\eta_{xx}=0 \label{eta}\end{equation} The general
solution of this equation is $\eta(x,t)=f(x-t)+g(x+t)$, where $f$
and $g$ are differentiable functions.  It is convenient first to
restrict ourselves to waves which propagate in only one direction,
thus, we choose
\begin{equation} \eta(x,t)=f(x-t) \label{sol}\end{equation}
From (\ref{4}), (\ref{sol}) and  the condition $v=\eta_t$ on
$z=1$, we obtain \be \mathcal{F}(x)=f(x)+c_0 \label{c}\ee where
$c_0$ is constant.\\
 Therefore, in the irrotational case, taking
into account (\ref{2}), (\ref{3''}), (\ref{4}), (\ref{sol}) and
(\ref{c}),  the solution of the linear system (\ref{small}) is
given by \be
 \begin{array}{llll}
 \eta(x,t)=f(x-t)\\
 p(x,t)=f(x-t)\\
u(x,z,t)=f(x-t)+c_0\\
v(x,z,t)=-zf'(x-t)=-zu_x \end{array}\label{solrot0}\ee

\section{Particles trajectories in the irrotational case}

Let $\left(x(t), z(t)\right)$ be the path of a particle in the
fluid domain, with location $\left(x(0), z(0)\right)$ at time
$t=0$. The motion of the particle is described by the differential
system \be
 \left\{\begin{array}{ll}
 \f{dx}{dt}=u(x,z,t)\\
 \f{dz}{dt}=v(x,z,t)
 \end{array}\right.\label{diff}\ee
with the initial data $\left(x(0), z(0)\right):=(x_0,z_0)$.

\noindent Making the \textit{Ansatz}

\be f(x-t)=\cos(2\pi(x-t))\label{ansatz} \ee from (\ref{solrot0}),
the differential system (\ref{diff}) becomes
\be\left\{\begin{array}{ll}
 \f{dx}{dt}=\cos(2\pi(x-t))+c_0\\
 \\
 \f{dz}{dt}=2\pi z\sin(2\pi(x-t))
 \end{array}\right.\label{diff2}\ee
 Notice that the constant $c_0$ is the average of the horizontal fluid
 velocity over any horizontal
 segment of length 1, that is,
 \be
 c_0=\f 1 {1}\int_{x}^{x+1}u(s,z,t)ds,
 \ee
 representing therefore the strength of the underlying uniform
 current. Thus, $c_0=0$  will correspond to a region of still water with
 no underlying current,
 $c_0>0$ will characterize a favorable uniform current and  $c_0<0$
 will characterize an adverse uniform current.

 The right-hand side of the differential system (\ref{diff2})
 is smooth and bounded, therefore, the unique solution of the Cauchy
 problem with initial data $(x_0,z_0)$ is defined globally in
 time.

 To study the exact solution of the system (\ref{diff2}) it is
 more convenient to re-write it in the following moving frame
 \be
 X=2\pi(x-t),\quad  Z=z \label{frame}
 \ee
This transformation yields \be\left\{\begin{array}{ll}
 \f{dX}{dt}=2\pi\cos(X)+2\pi(c_0-1)\\
 \\
 \f{dZ}{dt}=2\pi Z\sin(X)
 \end{array}\right.\label{diff3}\ee
Let us now investigate the differential system (\ref{diff3}).
\subsection{The case $\mathbf{c_0=0}$}

The  horizontal component of the velocity $u$ in (\ref{solrot0}),
with $f(x-t)$ given by (\ref{ansatz}), has in the moving frame
(\ref{frame}), the following expression \be u(X,Z,t)=\cos(X)+c_0
\ee Thus, the case $c_0=0$ is obtained for \be
\int_0^{2\pi}u(X,Z,t)dX=0 \ee This is the Stokes condition for
irrotational flows, that is, the horizontal velocity has a
vanishing mean over a period.\\
In the considered case, we write the first equation of the system
(\ref{diff3}) into the form \be \int \f{dX}{\cos(X)-1}=2\pi t
\label{kamke1}\ee We use the following substitution (see
\cite{kamke}, I.76, page 308) \be \sin(X)=\f{2y}{y^2+1}\, ,\quad
\cos(X)=\f{y^2-1}{y^2+1}\,, \quad
dX=-\f{2}{y^2+1}dy\label{substitution}\ee In the new variable,
(\ref{kamke1}) integrates at \be y=2\pi t+k \label{y}\ee k being
an integration constant. Hence, \be X(t)=2\textrm{arccot }(2\pi
t+k) \label{X}\ee Taking into account (\ref{substitution}),
(\ref{y}), we obtain \be \sin(X(t))=\f{2(2\pi t+k)}{1+(2\pi
t+k)^2} \ee

\noindent Therefore, the second equation in (\ref{diff3}) yields
\be Z(t)=Z(0)\exp \Bigg(\int_0^t2\pi \sin(X(s))\, ds\Bigg)
=Z(0)\exp\Bigg(\ln\Big[\f{1+(2\pi
t+k)^2}{1+k^2}\Big]\Bigg)\label{Z} \ee From (\ref{frame}),
(\ref{X}) and (\ref{Z}), we obtain that the solution of the system
(\ref{diff2}), with the initial data $(x_0,z_0)$, has the
following expression \be\left\{
\begin{array}{ll}x(t)=t+\f
1{\pi}\textrm{arccot }(2\pi t+k)\\
\cr z(t)=\f{z_0}{1+k^2}[1+(2\pi t+k)^2]
\end{array}\right.\label{solutie}\ee  From the initial conditions, we get $k:=\textrm{cot }(\pi
x_0)$. \\
The derivatives of $x(t)$ and $z(t)$ with respect to $t$, have the
expressions \begin{displaymath}
\begin{array}{ll}
x'(t)=\f{(2\pi t+k)^2-1}{1+(2\pi t+k)^2} \\ \cr
 z'(t)= \f{4\pi z_0}{1+k^2}(2\pi
t+k)
\end{array} \end{displaymath}
Therefore
\begin{displaymath}
\begin{array}{ll}
x'(t)>0 \, \Longleftrightarrow |2\pi t+k|>1 \\
 z'(t)>0\, \Longleftrightarrow (2\pi t+k)>0
\end{array} \end{displaymath}
the flat bottom being at $z=0$, we have $z_0>0$. \\
 Thus, for $t$ in the intervals $(-\infty,\,\f{-1-k}{2\pi})$,
$(\f{-1-k}{2\pi},\, -\f{k}{2\pi})$, $(-\f{k}{2\pi},
\,\f{1-k}{2\pi})$ and $(\f{1-k}{2\pi},\,\infty)$,
 the derivatives $x'(t)$, $z'(t)$, have the following signs
 \begin{eqnarray} t: && \quad \qquad \qquad \, \f{-1-k}{2\pi}
\quad \quad \, -\f{k}{2\pi} \quad \quad \,
\f{1-k}{2\pi}\nonumber\\
 &&\begin{picture}(1,1)\put(0,1){\line(1,0){215}}\end{picture} \label{derivate'}\end{eqnarray}
\vspace{-1.2cm}  \begin{eqnarray} \hspace{1.9cm}
\begin{array}{c|c|c|c}
 & & & \nonumber\\
 x'(t)>0 & x'(t)<0 & x'(t)<0
&
x'(t)>0\\
z'(t)<0 & z'(t)<0 & z'(t)>0 & z'(t)>0
\end{array}\nonumber\end{eqnarray}
The limits of $x(t)$, $z(t)$ and $\f{z(t)}{x(t)}$ for
$t\rightarrow -\infty$ and $t\rightarrow \infty$ are
\begin{eqnarray}
&&\lim_{t\rightarrow -\infty}x(t)=-\infty,\quad \lim_{t\rightarrow
\infty}x(t)=\infty
\nonumber\\
&&\lim_{t\rightarrow -\infty}z(t)=\infty,\quad \lim_{t\rightarrow
\infty}z(t)=\infty \label{limit}\\
&&\lim_{t\rightarrow -\infty}\f{z(t)}{x(t)}=-\infty,\quad
\lim_{t\rightarrow \infty}\f{z(t)}{x(t)}=\infty\nonumber
\end{eqnarray}
 Thus,
taking into account  (\ref{derivate'}) and (\ref{limit}), we
sketch below the graph of the parametric curve (\ref{solutie})

\vspace{0.6cm}

\hspace{2cm}\scalebox{0.70}{\includegraphics{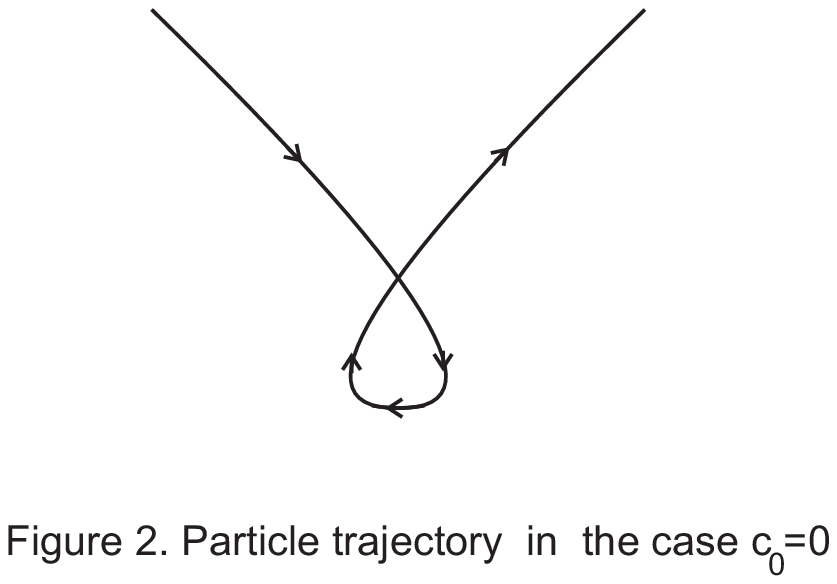}}
\\

\noindent Thus, we get:


\begin{theorem}

In the case of no underlying current, the particle trajectories
beneath the irrotational shallow water waves
 are curves which  have only one loop like in Figure 2.

\end{theorem}

\subsection{The case $\mathbf{c_0(c_0-2)>0}$} In this case, we
write the first equation of the system (\ref{diff3}) into the form
\be \int \f{dX}{\cos(X)+c_0-1}=2\pi t \label{kamke2}\ee We use the
same  substitution (\ref{substitution}). In the new variable,
(\ref{kamke2}) becomes \be -\f{2}{c_0}\int
\f{dy}{y^2+\f{c_0-2}{c_0}}=2\pi t \ee which integrates at \be
-\f{2}{c_0}\sqrt{\f{c_0}{c_0-2}}\arctan
\Big(\sqrt{\f{c_0}{c_0-2}}\, y\Big)=2\pi t+k \ee k being an
integration constant. Further, we obtain \be
y=\mathfrak{C}_0\tan(\alpha(t)), \label{Y}\ee
 with
\be \mathfrak{C}_0:=\sqrt{\f{c_0-2}{c_0}} \ee \be
\alpha(t):=-\f{c_0\mathfrak{C}_0}{2}(2\pi t+k)\ee Hence, returning
to the variable $X$, we get \be X(t)=2\textrm{arccot
}\Big[\mathfrak{C}_0\tan\left(\alpha(t)\right)\Big] \label{X1}\ee
 Taking into account (\ref{substitution}), (\ref{Y}), we obtain \be
\sin(X(t))=\f{2\mathfrak{C}_0\tan\left(\alpha(t)\right)}{1+\Big[\mathfrak{C}_0\tan\left(\alpha(t)\right)\Big]^2}
\ee
 The second equation
in (\ref{diff3}) yields \begin{eqnarray}
Z(t)=Z(0)\exp\Bigg(\int_0^t 2\pi\sin(X(s))ds\Bigg) \label{Z'}
\end{eqnarray} From (\ref{frame}), (\ref{X1}) and (\ref{Z'}), we
obtain that the solution of the system (\ref{diff2}), with the
initial data $(x_0,z_0)$, $z_0>0$, has the following expression
\be\left\{
\begin{array}{ll}x(t)=t+\f
1{\pi}\textrm{arccot
}\Big[\mathfrak{C}_0\tan\left(\alpha(t)\right)\Big]
\\
\cr z(t)=z_0 \exp\Bigg( \int_0^t
\f{4\pi\mathfrak{C}_0\tan\left(\alpha(s)\right)}{1+\Big[\mathfrak{C}_0\tan
\left(\alpha(s))\right)\Big]^2} ds\Bigg)\end{array}\right.\ee The
derivatives of $x(t)$ and $z(t)$ with respect to $t$, have the
expressions \begin{eqnarray}
\begin{array}{ll}
x'(t)=\f{
\left(\mathfrak{C}_0^2-1\right)\sin^2\left(\alpha(t)\right)+c_0-1}{\cos^2
\left(\alpha(t)\right)\left[1+\mathfrak{C}_0^2\tan^2\left(\alpha(t)\right)\right]}
=\left[\f{
2\Big(\sin^2\left(\alpha(t)\right)-\f{c_0^2-c_0}{2}\Big)}{(-c_0)}\right]\cdot\f{1}{\cos^2
\left(\alpha(t)\right)\left[1+\mathfrak{C}_0^2\tan^2\left(\alpha(t)\right)\right]}
\\ \cr
 z'(t)=z_0
\f{4\pi\mathfrak{C}_0\tan\left(\alpha(t)\right)}
{1+\Big[\mathfrak{C}_0\tan\left(\alpha(t)\right)\Big]^2}
 \exp\Bigg( \int_0^t
\f{4\pi\Big[\mathfrak{C}_0\tan\left(\alpha(s)\right)\Big]}{1+\Big[\mathfrak{C}_0\tan
\left(\alpha(s))\right)\Big]^2} ds\Bigg)
\end{array}\label{x'z'} \end{eqnarray}
Let us now  study the signs of the derivatives in (\ref{x'z'}). We
are in the case $\mathbf{c_0(c_0-2)>0}$, that is, $c_0\in
(-\infty,\, 0)\cup (2,\,\infty)$.

 \textbf{(a)} If
$\mathbf{c_0<-1}$, then $\f{c_0^2-c_0}{2}>1$. Therefore,
$\sin^2(\alpha(t))-\f{c_0^2-c_0}{2}<0$. Thus,  we obtain
 that $x'(t)<0$, for all $t$.\\
The sign of $z'(t)$ will depend on  $\alpha(t)$. For $\alpha(t)$
in intervals of the form
$\alpha(t)\in\left(-\f{\pi}{2}+l\pi,l\pi\right)$,
$l\in\mathbb{Z}$, we get $z'(t)<0$, and for
$\alpha(t)\in\left(l\pi,l\pi+\f{\pi}{2}\right)$, $l\in\mathbb{Z}$,
we get $z'(t)>0.$

We sketch below the particle trajectory in this case:

\vspace{1cm}

\hspace{2cm}\scalebox{0.70}{\includegraphics{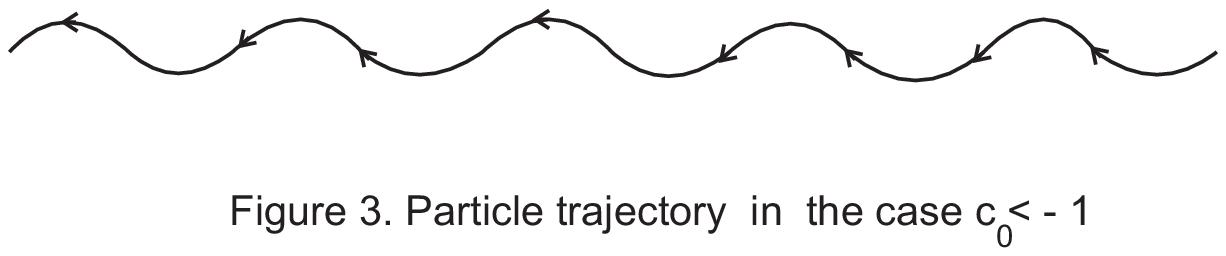}}
\\

 \textbf{(b)} If
$\mathbf{-1\leq c_0<0}$, then $\f{c_0^2-c_0}{2}\leq 1$.
Thus, \\
 for $\alpha(t)<-\arcsin\left(\sqrt{\f{c_0^2-c_0}{2}}\right)+l\pi$,
 we get $x'(t)>0$, $z'(t)<0$,\\
for $\alpha(t)\in
\Bigg(-\arcsin\left(\sqrt{\f{c_0^2-c_0}{2}}\right)+l\pi,\,
l\pi\Bigg)$,
we get  $x'(t)<0$, $z'(t)<0$,\\
 for  $\alpha(t)\in \Bigg(l\pi,
\,\arcsin\left(\sqrt{\f{c_0^2-c_0}{2}}\right)+l\pi\Bigg)$
we get  $x'(t)<0$, $z'(t)>0$,\\
for $\alpha(t)>\arcsin\left(\sqrt{\f{c_0^2-c_0}{2}}\right)+l\pi$,
we get $x'(t)>0$, $z'(t)>0$,\\
where $l\in \mathbb{Z}$. We sketch below the particle trajectory
in this case:

\vspace{1cm}

\hspace{2cm}\scalebox{0.70}{\includegraphics{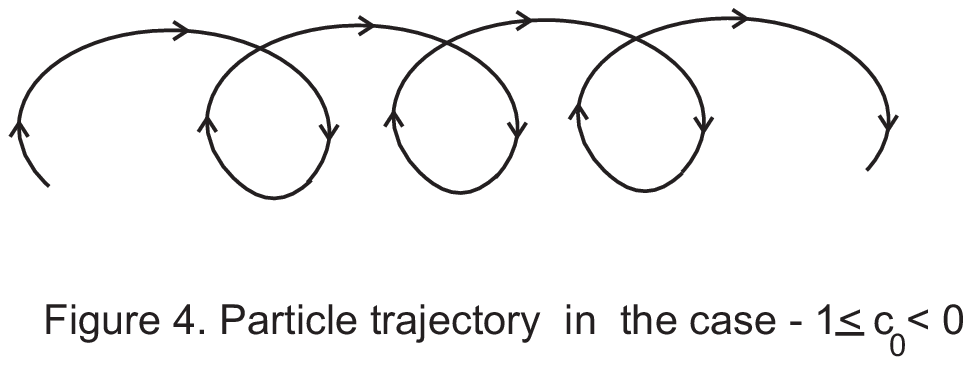}}
\\

 \textbf{(c)} If
$\mathbf{c_0>2}$, then $\f{c_0^2-c_0}{2}>1$. Therefore,
$\sin^2(\alpha(t))-\f{c_0^2-c_0}{2}<0$.  Thus,  we obtain
 that $x'(t)>0$, for all $t$.\\
 The sign of $z'(t)$ will depend on  $\alpha(t)$.
$z'(t)<0$ for $\alpha(t)\in\left(-\f{\pi}{2}+l\pi,l\pi\right)$,
$l\in\mathbb{Z}$, and $z'(t)>0$ for
$\alpha(t)\in\left(l\pi,l\pi+\f{\pi}{2}\right)$, $l\in\mathbb{Z}$.

We sketch below the particle trajectory in this case:

\vspace{1cm}

\hspace{2cm}\scalebox{0.70}{\includegraphics{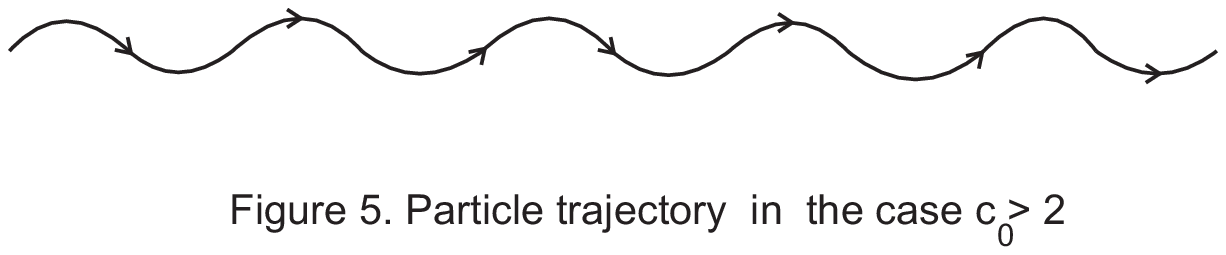}}
\\

\noindent Therefore, we proved:

\begin{theorem}

 In the case that the underlying uniform current is moving in the
same direction as an irrotational shallow water wave and the
strength of the current is bigger than 2, then  the particles
trajectories beneath the wave are undulating paths to the right
(see Figure 5).

 In the case that the underlying uniform current is moving in the
opposite  direction as an irrotational shallow water wave and the
strength of the current is smaller than -1, then  the particles
trajectories beneath the wave are undulating paths to the left
(see  Figure 3). If the strength of the adverse current is bigger
than -1, then the particle trajectories are loops with positive
drift (see Figure 4).

\end{theorem}

\subsection{The case $\mathbf{c_0\in (0,2]}$} In this case, we
write the first equation of the system (\ref{diff3}) into the form
\be \int \f{dX}{\cos(X)+c_0-1}=2\pi t \label{kamke2'}\ee We use
the same  substitution (\ref{substitution}). In the new variable,
(\ref{kamke2'}) becomes \be -\f{2}{c_0}\int
\f{dy}{y^2-\f{2-c_0}{c_0}}=2\pi t \ee which integrates at \be
-\f{1}{c_0}\sqrt{\f{c_0}{2-c_0}}\ln
\Bigg|\f{y-\sqrt{\f{2-c_0}{c_0}}}{y+\sqrt{\f{2-c_0}{c_0}}}\Bigg|=2\pi
t+k \ee k being an integration constant. Further, we obtain
\begin{eqnarray}
y=\mathfrak{K}_0\f{\exp(2\beta(t))+1}{\exp(2\beta(t))-1} \, \,
\textrm{ if } \, \, |y|>\mathfrak{K}_0\, ,\nonumber \\
y=\mathfrak{K}_0\f{\exp(2\beta(t))-1}{\exp(2\beta(t))+1} \, \,
\textrm{ if } \, \, |y|<\mathfrak{K}_0 \,
,\label{Y1}\end{eqnarray} where \be
\mathfrak{K}_0:=\sqrt{\f{2-c_0}{c_0}} \ee \be
\beta(t):=\f{c_0\mathfrak{K}_0}{2}(2\pi t+k)\ee Hence, returning
to the variable $X$, we get \be X(t)=2\textrm{arccot
}\Big[\mathfrak{K}_0\coth\left(\beta(t)\right)\Big] \, \, \textrm{
or } \, \, X(t)=2\textrm{arccot
}\Big[\mathfrak{K}_0\tanh\left(\beta(t)\right)\Big]\label{X1'}\ee
if $|\cot\left(\f{X}{2}\right)|>\mathfrak{K}_0$, respectively,
$|\cot\left(\f{X}{2}\right)|<\mathfrak{K}_0$.\\
 Taking into account (\ref{substitution}), (\ref{Y1}), we obtain \be
\sin(X(t))=\f{2\mathfrak{K}_0\coth\left(\beta(t)\right)}{1+\Big[\mathfrak{K}_0
\coth\left(\beta(t)\right)\Big]^2} \, \, \textrm{ or } \, \,
\sin(X(t))=\f{2\mathfrak{K}_0\tanh\left(\beta(t)\right)}{1+\Big[\mathfrak{K}_0
\tanh\left(\beta(t)\right)\Big]^2}\ee Thus, the solution of the
system (\ref{diff2}), with the initial data $(x_0,z_0)$, $z_0>0$,
has in this case the following expressions \be\left\{
\begin{array}{ll}x(t)=t+\f
1{\pi}\textrm{arccot
}\Big[\mathfrak{K}_0\coth\left(\beta(t)\right)\Big]
\\
\cr z(t)=z_0 \exp\Bigg( \int_0^t
\f{4\pi\mathfrak{K}_0\coth\left(\beta(s)\right)}{1+\Big[\mathfrak{K}_0\coth
\left(\beta(s))\right)\Big]^2} ds\Bigg)
\end{array}\right.\label{sol1}\ee or \be\left\{
\begin{array}{ll}x(t)=t+\f
1{\pi}\textrm{arccot
}\Big[\mathfrak{K}_0\tanh\left(\beta(t)\right)\Big]
\\
\cr z(t)=z_0 \exp\Bigg( \int_0^t
\f{4\pi\mathfrak{K}_0\tanh\left(\beta(s)\right)}{1+\Big[\mathfrak{K}_0\tanh
\left(\beta(s))\right)\Big]^2} ds\Bigg)
\end{array}\right.\label{sol2}\ee We derive  $x(t)$ and $z(t)$ from
(\ref{sol1})  with respect to $t$ and we get \be\left\{
\begin{array}{ll}x'(t)=1+\f{2-c_0}{
\sinh^2\left(\beta(t)\right)\Big[1+\mathfrak{K}_0^2\coth^2\left(\beta(t)\right)\Big]}
\\
\cr
z'(t)=z_0\f{4\pi\mathfrak{K}_0\coth\left(\beta(t)\right)}{1+\Big[\mathfrak{K}_0\coth
\left(\beta(t))\right)\Big]^2} \exp\Bigg( \int_0^t
\f{4\pi\mathfrak{K}_0\coth\left(\beta(s)\right)}{1+\Big[\mathfrak{K}_0\coth
\left(\beta(s))\right)\Big]^2} ds\Bigg)
\end{array}\right.\ee
Because we are in the case $c_0\in (0,2]$, we have $2-c_0>0$.
Thus, the derivative $x'(t)>0$ for all $t$. The sign of $z'(t)$
depends on the sign of $\beta(t)$, that is, for $\beta(t)<0$ we
have $z'(t)<0$ and for $\beta(t)>0$ we have $z'(t)>0$. Then, the
particle trajectory in this case is like in Figure 6 (a).

For the second alternative  (\ref{sol2}), we get
 \be\left\{
\begin{array}{ll}x'(t)=\f{\left(\mathfrak{K}_0^2+1\right)\sinh^2\left(\beta(t)\right)+c_0-1
}{\cosh^2\left(\beta(t)\right)\left[1+\mathfrak{K}_0^2\tanh^2\left(\beta(t)\right)\right]}
=\left[\f{2\left(\sinh^2\left(\beta(t)\right)-\f{c_0-c_0^2}{2}\right)}{c_0}\right]\f{1}{\cosh^2\left(\beta(t)\right)\left[1+\mathfrak{K}_0^2\tanh^2\left(\beta(t)\right)\right]}
\\
\cr z'(t)=z_0
\f{4\pi\mathfrak{K}_0\tanh\left(\beta(s)\right)}{1+\Big[\mathfrak{K}_0\tanh
\left(\beta(s))\right)\Big]^2} \exp\Bigg( \int_0^t
\f{4\pi\mathfrak{K}_0\tanh\left(\beta(s)\right)}{1+\Big[\mathfrak{K}_0\tanh
\left(\beta(s))\right)\Big]^2} ds\Bigg) \end{array}\right.\ee
 \textbf{(a)} If
$\mathbf{1<c_0\leq 2}$, then
 we get $x'(t)>0$, for all $t$.\\
The sign of $z'(t)$ will depend on  $\beta(t)$. For $\beta(t)<0$,
we get $z'(t)<0$, and for $\beta(t)>0$, we get $z'(t)>0.$\\
The particle trajectory in this case  is like in Figure 6 (a).
\\

 \textbf{(b)} If
$\mathbf{0<c_0\leq 1}$, then \\
 for $\sinh(\beta(t))<-\sqrt{\f{c_0-c_0^2}{2}}$, we get $x'(t)>0$, $z'(t)<0$,\\
for $\sinh(\beta(t))\in \Bigg(-\sqrt{\f{c_0-c_0^2}{2}},\,
0\Bigg)$,
we get  $x'(t)<0$, $z'(t)<0$,\\
 for  $\sinh(\beta(t))\in \Bigg(0,
\,\sqrt{\f{c_0-c_0^2}{2}}\Bigg)$
we get  $x'(t)<0$, $z'(t)>0$,\\
for $\sinh(\beta(t))>\sqrt{\f{c_0-c_0^2}{2}}$,
we get $x'(t)>0$, $z'(t)>0$,\\
Thus, the particle  trajectory in this case is sketched  in Figure
6 (b).

 \vspace{0.7cm}

\hspace{2cm}\scalebox{0.70}{\includegraphics{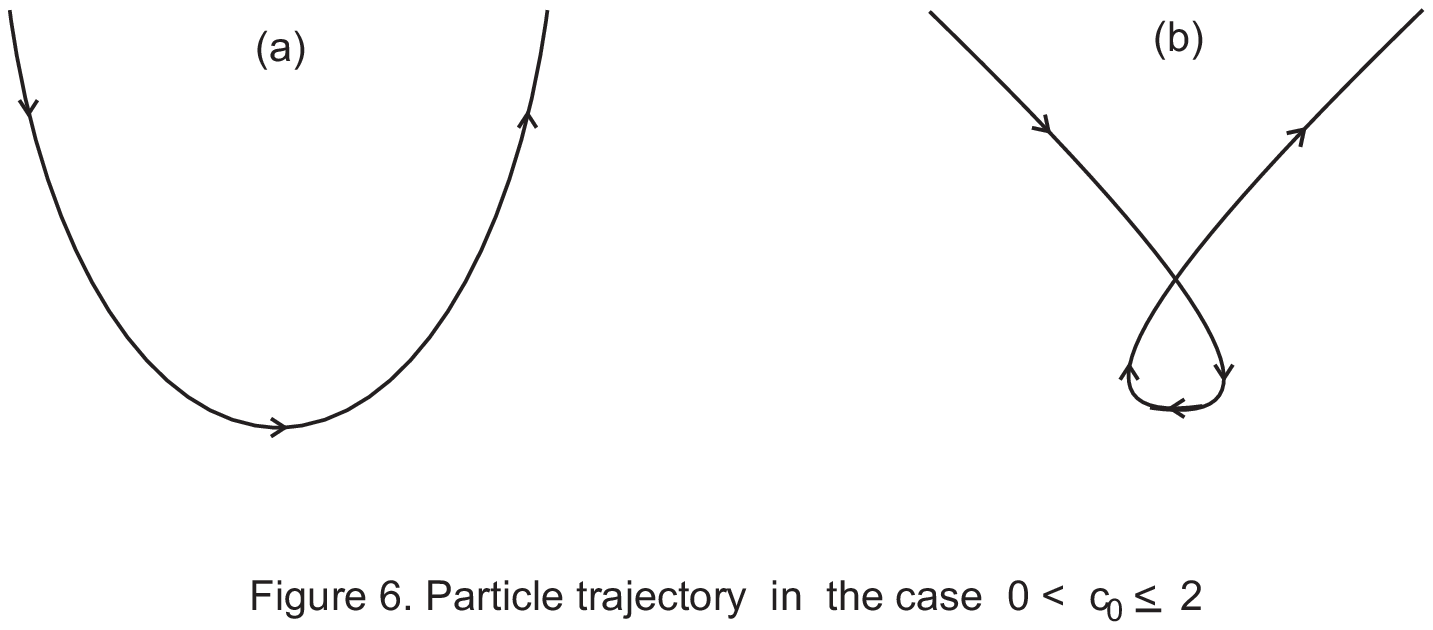}}
\\

\noindent We thus have:

\begin{theorem}
 In the case that the underlying uniform current is moving in the
same direction as an irrotational shallow water wave and the
strength of the current is smaller than 2, then  the particles
trajectories beneath the wave are parabolic curves or curves which
have only one loop like in Figure 6.
\end{theorem}

\label{lastpage}


\begin{thebibliography}{99}

\bibitem{aft} \textsc{Amick C. J.}, \textsc{Fraenkel L. E.},
\textsc{Toland J. F.}, On the Stokes conjecture for the wave of
extreme form, \textit{Acta Math.} \textbf{148} (1982), 193--214.
\bibitem{c2001a} \textsc{Constantin A.}, On the deep water wave
motion, \textit{J. Phys. A} \textbf{34} (2001), 1405--1417.
\bibitem{c2001b} \textsc{Constantin A.}, Edge waves along  a sloping
beach, \textit{J. Phys. A} \textbf{34} (2001), 9723--9731.
\bibitem{c2007} \textsc{Constantin A.},
The trajectories of particles in Stokes waves, \textit{Invent.
Math.}
 \textbf{166} (2006), 523--535.
 \bibitem{cev} \textsc{Constantin A.}, \textsc{Ehrnstr\"{o}m M.}, \textsc{Villari G.},
Particle trajectories in linear deep-water waves,
\textit{Nonlinear Anal. Real World Appl.},
doi:10.1016/j.nonrwa.2007.03.003.
\bibitem{CE2} \textsc{Constantin A.}, \textsc{Escher J.},
Particle trajectories in solitary water waves, \textit{ Bull.
Amer. Math. Soc.} \textbf{44} (2007), 423--431.
\bibitem{cs2'} \textsc{Constantin A.},  \textsc{Strauss W.},
Exact steady periodic water waves with vorticity, \textit{Comm.
Pure Appl. Math.} \textbf{57} (2004), 481--527.
\bibitem{cs2007} \textsc{Constantin A.},  \textsc{Strauss W.},
Stability properties of steady water waves with vorticity,
\textit{Comm. Pure Appl. Math.} \textbf{60} (2007), 911--950.
\bibitem{CS} \textsc{Constantin A.},  \textsc{Strauss W.},
Pressure and trajectories beneath a Stokes wave, Preprint (2008).
\bibitem{cv} \textsc{Constantin A.}, \textsc{Villari G.},
Particle trajectories in linear water waves, \textit{J. Math.
Fluid Mech.} \textbf{10} (2008), 1--18.
\bibitem{shkoller} \textsc{Coutand D.}, \textsc{Shkoller S.},
Well-posedness of the free-surface incompressible Euler equations
with or without surface tension, \textit{J. Amer. Math. Soc.}
\textbf{20} (2007), 829-930.
\bibitem{crapper} \textsc{Crapper G. D.}, An exact solution for
progressive capillary waves of arbitrary amplitude, \textit{J.
Fluid Mech.} \textbf{2} (1957), 532--540.
\bibitem{debnath} \textsc{Debnath L.}, Nonlinear Water Waves,
Boston, MA: Academic Press Inc., 1994.
\bibitem{gerstner} \textsc{Gerstner F.}, Theorie der Wellen samt
einer daraus abgeleiteten Theorie der Deichprofile, \textit{Ann.
Phys.} \textbf{2} (1809), 412--445.
\bibitem{henry} \textsc{Henry D.}, The trajectories of particles
in deep-water Stokes waves, \textit{Int. Math. Res. Not.} (2006),
Art. ID 23405, 13 pp.
\bibitem{johnson-carte} \textsc{Johnson R. S.},
A Modern Introduction to the Mathematical Theory of Water Waves,
Cambridge Univeristy Press, 1997.
\bibitem{kamke} \textsc{ Kamke E.},  Differentialgleichungen, L\"{o}sungsmethoden
und L\"{o}sungen, vol. I,  Akademische Verlagsgesellschaft Geest
\& Portig K.-G., Leipzig, 1967.
\bibitem{kk} \textsc{Kenyon K. E.}, Shallow water gravity waves: a
note on the particle orbits, \textit{J. Oceanography} \textbf{52}
(1996), 353--357.
\bibitem{kinn} \textsc{Kinnersley W.}, Exact large amplitude capillary waves
on sheets of fluids, \textit{J. Fluid Mech.} \textbf{77} (1976),
229--241.
\bibitem{lamb} \textsc{Lamb H.} Hydrodynamics (Sixth Edition),
Dover Publications, New York, 1945.
\bibitem{lighthill} \textsc{Lighthill J.}, Waves in Fluids,
Cambridge University Press, 2001.
\bibitem{sommerfeld} \textsc{Sommerfeld A.}, Mechanics of
Deformable Bodies, New York: Academic Press Inc., 1950.
\bibitem{stoker} \textsc{Stoker J. J.}, Water Waves. The
Mathematical Theory with Applications, New York: Interscience
Publ. Inc., 1957.
\bibitem{stokes} \textsc{Stokes G. G.}, On the theory of
oscillatory waves, \textit{Trans. Camb. Phil. Soc.} \textbf{8}
(1847), 441--455. Reprinted in: \textsc{Stokes G. G.},
Mathematical and Physical Papers, Volume I. Cambridge University
Press, 197--229, 1880.
\bibitem{toland} \textsc{Toland J. F.},
Stokes waves, \textit{Topol. Methods Nonlinear Anal.} \textbf{7}
(1996), 1--48.


\end{thebibliography}
\end{document}